\documentclass{PoS}

\title{Wind accretion in symbiotic X-ray binaries}

\ShortTitle{Wind accretion in SyXBs}

\author{\speaker{K.~Postnov}$^a$, N.~Shakura$^a$, A.~Gonz\'alez-Gal\'an$^b$, E.~Kuulkers$^c$, P.~Kretschmar$^c$, S.~Larsson$^d$, M.H.~Finger$^{e,f}$, A.~Kochetkova$^a$, G. L\"u$^g$ and L.~Yungelson$^h$ \\
    \llap{$^a$}Sternberg Astronomical Institute, 13, Universitetskij pr., 119992 Moscow, Russia\\
\llap{$^b$} Departamento de F\'isica, Ingenier\'ia de Sistemas y Teor\'ia de la Se\~nal,Universidad de Alicante, Apdo. 99, 03080 Alicante, Spain\\
\llap{$^c$} European Space Astronomy Centre (ESAC), P.O.~Box 78, 28691, Villanueva de la Ca\~nada, Spain\\
\llap{$^d$} Department of Astronomy, Stockholm University, SE-106 91 Stockholm, Sweden\\
\llap{$^e$} National Space Science and Technology Center, 320 Sparkman Drive, Huntsville, AL 35805, USA\\
\llap{$^f$} Universities Space Research Association, 6767 Old Madison Pike, Suite 450, Huntsville, AL 35806, USA\\
 \llap{$^g$} School of Physics, Xinjiang University, Urumqi, 830046 China\\
\llap{$^h$} Institute of Astronomy RAS, Moscow, 48
Pyatnitskaya Str., Moscow, 119017, Russia\\
        E-mail: \email{kpostnov@gmail.com}, \email{nikolai.shakura@gmail.com},  \email{anagonzalez@ua.es},\email{Erik.Kuulkers@sciops.esa.int} \email{Peter.Kretschmar@esa.int}, \email{stefan@astro.su.se}, \email{mark.finger@nasa.gov}, \email{sunny@sai.msu.ru}, 
\email{guolianglv@sina.com}, \email{lry@inasan.ru}}

\abstract{The properties of wind accretion in symbiotic X-ray binaries (SyXBs) 
consisting of red-giant and magnetized neutron star (NS) are discussed. 
The spin-up/spin-down torques applied to NS are derived based on a hydrodynamic
theory of quasi-spherical accretion onto magnetized NSs.
In this model, a settling subsonic accretion 
proceeds through a hot shell formed around the NS magnetosphere.
The accretion rate onto the NS is determined by the ability of the plasma to enter the
magnetosphere. 
Due to large Reynolds numbers in the shell, 
the interaction of the rotating magnetosphere with plasma 
initiates a subsonic turbulence. The convective motions
are capable of carrying the angular momentum through the shell. 
We carry out a population synthesis of SyXBs in the Galaxy with account 
for the spin evolution of magnetized NS. 
The Galactic number of SyXBs with bright (M$_{v}<1$) low-mass red-giant companion 
is found to be from $\sim$ 40
to 120, and their birthrate is $\sim 5\times
10^{-5}-10^{-4}$ yr$^{-1}$. 
According to our model,
among known SyXBs, Sct X-1 and IRXS J180431.1-273932 are wind-fed
accretors. GX 1+4 lies in the transition from the wind-fed SyXBs to
SyXBs in which the giants overflow their Roche lobe.  
The model successfully reproduces very long NS spins 
(such as in IGR J16358-4724 and 4U 1954+31) without the need to invoke
very strong magnetic fields.
}

\FullConference{8th INTEGRAL Workshop ''The Restless Gamma-ray Universe''--Integral2010,\\
		September 27-30, 2010\\
		Dublin Ireland}

\begin{document}

\section{Introduction}

\begin{table*}
{\footnotesize
  \tabcolsep0.70mm
 \begin{tabular}{lccccl}
  \hline\hline
  SyXB&$P_{\rm s}$ (s)&$\dot{P_{\rm s}}/P_{\rm s}$&$P_{\rm orb}$ (d)&$L_{\rm
  X}$ (erg s$^{-1}$)&Sp. Type\\
  \hline
 GX 1+4 &150$^{\cite{cr97}}$& $\pm 4\times 10^{-10}$&1161$^{\cite{hi06}}$
 &$\sim 10^{37}$$^{\cite{cr97}}$&M5 III$^{\cite{cr97}}$\\
 4U 1954+31&$\sim 18300$$^{\cite{corb08}}$&$-1.4\times10^{-9}$$^{\cite{corb08}}$&?
 &$4\times10^{32}-10^{35}$$^{\cite{ma06}}$&M4 III$^{\cite{ma06}}$\\
4U 1700+24&?&?&404$^{\cite{ma02}}$&$2\times10^{32}-10^{34}$$^{\cite{ka07}}$&M2 III$^{\cite{ma02}}$\\
Sct X-1&113$^{\cite{ka07}}$&$3.9\times10^{-9}$$^{\cite{ka07}}$&?&$2\times10^{34}$$^{\cite{ka07}}$
&Late K/early M I-III$^{\cite{ka07}}$\\
IGR J16194-2810 &?&?&?&$\leq 7\times10^{34}$$^{\cite{ma07}}$&M2 III$^{\cite{ma07}}$\\
IRXS J180431.1-273932&494$^{\cite{nu07}}$&?&?&$\leq6\times10^{34}$$^{\cite{nu07}}$
&M6 III$^{\cite{nu07}}$\\
IGR J16358-4724&5850$^{\cite{pate04}}$&$3.1\times10^{-8}$$^{\cite{pate07}}$&?&$3\times10^{32}-3\times10^{36}$$^{\cite{pate07}}$ &K-M III$^{\cite{nesp09}}$\\
IGR J16393-4643&912$^{\cite{boda06, thom06}}$&$1.0\times10^{-11}$$^{\cite{nesp09}}$&50.2$^{\cite{nesp09}}$&?
&K-M III$^{\cite{nesp09}}$ \\
2XMM J174016.0-290337&626$^{\cite{far10}}$&?&?&$\sim3\times10^{34}$$^{\cite{far10}}$&K1 III$^{\cite{far10}}$ \\
\label{tab:syxb}
\end{tabular}
\caption{Parameters of observed symbiotic X-ray binaries}
}
\end{table*}

Symbiotic X-ray binaries (SyXB) form a growing subclass of 
X-ray binary stars in which 
accretion onto a magnetized neutron star (NS) occurs 
from stellar wind of the companion red giant 
that does not fill its Roche lobe \cite{ma06}. All these systems are hard X-ray emitters
\cite{mu97} and  are low-mass X-ray binaries, with the possible
exception of Sct X-1 \cite{ka07}, for which a high-mass solution
also is viable. One of these systems,  IGR J16358-4726 may harbour  a magnetar  \cite{pate07}. 
Currently about ten sources are classified as  SyXBs or  candidate systems
(see   Table \ref{tab:syxb}).

Remarkable features
of SyXBs compared to other low-mass
X-ray binaries are  their long orbital periods and long NS spin periods.
In the catalog of 187 low-mass X-ray binaries \cite{liu07},
the orbital periods of 4U 1700+24 and GX 1+4 are the longest in
all orbital periods observed. With  18300\,s spin period,  4U~1954+31 is
the slowest accretion powered NS \cite{ma06}.
Therefore, a detailed investigation of SyXB is of immense importance for comprehension of
evolution of NS in LMXB.

\section{Quasi-spherical wind accretion}

The cold stellar wind of the red-giant companion 
is gravitationally captured by NS at the characteristic distance 
$R_G=2GM/(V_w^2+v_{orb}^2)^2$ (the Bondi radius). 
Here $V_w$ is the stellar wind velocity at the NS orbital distance, 
$v_{orb}$ is the NS orbital velocity. We consider the case of ineffective
cooling of plasma behind the shock, so the bow shock is formed at the
characteristic radius $\sim R_G$.
Behind the shock front, the matter
falls with a subsonic velocity  toward the NS magnetosphere located at a distance $R_A$
(the Alfven radius) which is determined by the pressure balance in the accreting matter and 
the magnetic field. To within a numerical factor of the order of a few, which depends on
the magnetosphere structure,  $R_A\sim [\mu^2/(\dot M\sqrt{2GM})]^{2/7}$. 

The matter carries the specific angular momentum $j_w=k_w\Omega_B R_G^2$ 
(here $-1<k_w<0.4$ is the numerical coefficient \cite{Ho_ea}, 
$\Omega_B$ is the binary orbital angular frequency). If $j_w>j_K(R_A)=\sqrt{GMR_A}$, 
an accretion disc is formed around the NS magnetosphere. In the opposite case, 
the accretion proceeds quasi-spherically. The interaction of accreting 
matter with the NS magnetosphere ultimately leads to the NS spin-up or spin-down -- the phenomenon 
that has been intensively studied \cite{Bildsten_ea97}.  

The disc accretion on NS magnetospheres is well studied
(see \cite{Lovelace_ea95} for the discussion) and can explain
many properties of accreting X-ray pulsars. However, 
the analysis of spin-down torques in a prototypical SyXB GX 1+4 \cite{Chakrabarty_ea97, Gonzalez_Galan_ea10} 
shows that the disc accretion can hardly explain the observed correlations.    

The quasi-spherical accretion 
onto slowly rotating magnetized NSs has recently been revised in \cite{Shakura_ea10}. 
In contrast to Bondi accretion, if the cooling is ineffective,
the matter can settle onto the NS magnetosphere with a subsonic velocity
through a quasi-spherical hot shell. 
The structure of this shell can be found from the hydrostatic equilibrium: 
$\rho(R)\sim {R}^{-3/2}$, $P(R)\sim {R}^{-5/2}$, $T(R)\sim R^{-1}$. The hot 
plasma near the shell bottom can be marginally stable against interchange 
instabilitites mediating plasma entering the magnetosphere, and for accretion to 
become possible the plasma cooling (e.g. via Compton scattering of X-ray photons) 
may be required \cite{AronsLea76}. However, the very fact of observations of slow X-ray pulsars 
accreting from stellar wind and experiencing spin-down episodes 
suggests that plasma finds ways to enter the magnetosphere.

In our model the accretion rate onto the NS, which determines its X-ray
luminosity, 
is always smaller than the Bondi-Hoyle accretion rate potentially available 
by the neutron star $\dot M_G\sim \rho_w R_G^2/v_w^3$; 
the extra matter gravitationally captured by the NS 
should bend around the hot shell and carry away the angular momentum in
the accretion wake. 

As shown in 
\cite{Shakura_ea10}, the Reynolds number in the hot shell is very high so 
the interaction of the rotating magnetosphere with the shell 
initiates subsonic turbulence. At the bottom of the shell the 
energy is released due to interaction of the shell with the magnetosphere
and turbulence, so powerful convective motions begins and establish 
nearly iso-angular-momentum rotation law in the shell: $\omega(R)\sim R^{-2}$.  
The NS will spin-up or spin-down depending on the specific 
angular momentum of matter at the magnetospheric boundary and 
the angular velocity of the NS rotation $\omega^*=2\pi/P_s$. 
The spin-up/spin-down equation of the quasi-spherically accreting neutron star
can be written in the form:
\begin{equation}
\label{eq1}
I\dot \omega^*=Z \dot M \omega_B R_G^{2}-(Z-z)\dot M R_A^2\omega^*\,,
\end{equation}
where $Z$ and $0<z<1$ are some dimensionless coefficients.  
Clearly, for spin-down episodes to become possible the condition $Z>z$ must 
be satisfied. Analysis of torque reversals observed in wind-accreting X-ray pulsars 
GX 1+4, Vela X-1, GX 302-1 and Cen X-3 shows that in the case of quasi-spherical
accretion $Z\gtrsim 1$ in these pulsars \cite{Shakura_ea10}.
The above equation for the NS spin evolution allows us to estimate the NS magnetic moment 
of X-ray pulsars in which $\dot \omega^*$ can be measured as a function of the X-ray flux
near the torque reversal point (i.e. where $\dot \omega^*=0$).
Remarkably, the estimate of the NS magnetic moment $\mu$
is independent of the actual value of the X-ray luminosity, 
only $\dot M(\partial \dot\omega^*/\partial \dot M)$ near the torque reversal
point matters. 
The magnetic field of NS in Vela X-1 and GX 301-2 estimated in this way 
is consistent with the value inferred from the cyclotron line measurements. 

The very slow pulsar spins in SyXBs ($\omega^* R_A<\omega_K(R_A)$) make 
it hard to establish the propeller regime 
to eject matter with parabolic velocities from the magnetosphere during spin-down episodes.
The important 
ratio of viscose tensions to the gas pressure at the magnetospheric boundary 
is proportional to $(\omega^*-\omega(R_A))/\omega_K(R_A)$, with the maximum 
value for the  non-rotating envelope with 
$\omega(R_A)=0$. For very slowly rotating pulsars this ratio is 
very small, i.e. only large-scale convective 
motions with the characteristic hierarchy of 
eddies scaled with radius can be established in the shell. 
When $\omega^*>\omega_K(R_A)$, the super-sonic propeller regime
must set in. In that case the maximum possible spin-down torque is $\sim \mu^2/R_A^3$.
It is not excluded that a hot envelope will persist here, too, 
and bring away the angular momentum from the rotating magnetosphere. 
If the characteristic gas cooling time in the envelope is short, 
one can expect the formation of a 'storaging' thin Keplerian disc \cite{Shakura_Sunyaev76}. 
There is no accretion of matter through such a disc, it only serves to remove the angular momentum from the magnetosphere. 

At high $\dot M_G$, thermal instability can
appear in hot plasma behind the bow shock front. 
The medium can be stratified into cold dense and hot rarefied regions. Cold blobs move in the hot 
(${\cal R} T\sim GM/R$) plasma. As they move deeper into the hot shell, their velocity can exceed the 
local sound velocity, and blobs get gradually destroyed. Crossing the line of sight, these cold blobs 
would produce an X-ray flux variability with increased hardness ratio. 
At high X-ray luminosities (high $\dot M$) processes of Compton heating and cooling near and above
the Alfven radius 
become important. Below some characteristic distance 
the Compton cooling by X-rays dominates, and above it the Compton heating prevails. 
This radius can be determined from the condition $T(R_x)\simeq T_X$, where $T_X$ is the effective temperature of X-ray photons. The latter depends
on the spectral shape (for example,  $T_X=1/4 T$ for spectrum with an exponential cutoff 
$\simeq e^{-E/kT}$,  $T_X=T$ for the Planckian spectrum).
If the Compton cooling time is shorter than the fall time of matter, non-stationary accretion regime sets in \cite{Sunyaev78}.
Real wind-accreting systems demonstrate complex quasi-stationary behavior with dips, 
outbursts, etc. These features can be interpreted using the proposed model of quasi-spherical accretion.  

\section{Population synthesis of SyXB}

\begin{figure}
\includegraphics[height=0.49\textwidth, width=2in,angle=-90]{spev2.ps}
\hfill
\includegraphics[width=2in, height=0.49\textwidth, angle=-90]{psma.ps}
\parbox[t]{0.49\textwidth}{\caption{The spin evolution of an accreting magnetized NS in a binary
system}\label{fig:spev}}
\hfill
\parbox[t]{0.49\textwidth}{\caption{X-ray luminosities 
and pulsar spin periods in SyXBs with bright red-giants for different models
(see \cite{Lu_ea10} for more detail). 
Open circles show sources from Table 1.}\label{fig:psma}}
\end{figure}
For the simulation of binary evolution, we use the rapid binary star
evolution code BSE \cite{hu02} with updates by \cite{kiel06}.
The code is appended by an algorithm for the careful treatment of spin evolution of a
magnetized NS and the NS magnetic field decay, as described in \cite{Lu_ea10}. The important parameter 
for the SyXB evolution is the stellar wind velocity from the red giant,
which we parametrize as 
$v_w=\sqrt{\beta (2GM/R)}$. The parameter $\beta$ is set to 1.

An example of the spin evolution of a magnetized NS in SyXB is presented in Fig. \ref{fig:spev}.
The parameter for the specific angular momentum of captured 
stellar wind is $k_w=1/4$.
We start with a  binary consisting of a 1.4 $M_\odot$ NS with the surface 
magnetic field $5\times10^{12}$~G, the spin period 
$P_s=0.01$~s, and a 1.3 $M_\odot$ main-sequence star. 
The initial orbital period of the system is $P_B=400$~d. 
Before point
'A', the secondary is still a MS-star and $\dot M_G$ onto the NS
is below $10^{-15} M_\odot$ yr$^{-1}$, so the NS evolves like an isolated
object. The evolution of $P_{\rm s}$ is determined by the spin-down
torque calculated by the magneto-dipole formula.
At point 'A' the
secondary enters the Hertzsprung gap and starts to lose mass
via stellar wind. The Bondi radius is $R_{\rm G}=1.3\times
10^{13} {\rm cm}$, relations  $R_G>R_A>R_c=(GM/\omega^{*2})^{1/3}$  and
$j_w<j_K(R_{A})$ hold, so the NS enters the super-sonic propeller stage
of quasi-spherical accretion. The NS spin period (and hence the corotation radius $R_c$)
increases, and 
at point 'B' $R_A=R_c$. The NS turns out at the quasi-spherical 
accretor state from stellar wind, and
the evolution of $P_{s}$ is determined by
Eq. \ref{eq1}. The SyXB stage starts.
At point 'C', the secondary evolves to the first giant branch (FGB)
stage. The NS is still in the accretor state. With the ascend along FGB,
the stellar wind velocity of the giant $v_w$ decreases so that 
$j_w\sim v_w^{-4}$ increases. At point 'D', $j_w>j_K(R_{A})$, 
and a wind-fed accretion disc forms around the NS. 
At point 'E' the secondary overfills its Roche lobe, the mass accretion
rate immediately increases and exceeds the critical Eddington rate. 
The NS becomes a super-accretor.
At point 'F', the secondary leaves FGB and evolves to the core
helium burning stage. The mass-loss rate rapidly decreases, but the
velocity of the stellar wind is low so that $j_w>j_{K}(R_{A})$ 
and the NS becomes a propeller with a wind-fed accretion disc.
At point 'G', the secondary evolves to early asymptotic giant branch
(E-AGB). At point 'H' $R_{A}\le R_{c}$ and the NS becomes a 
wind-fed accretor again. For a short time the system again appears as a SyXB. 
Due to the high mass accretion rate, the spin period $P_{\rm s}$ rapidly
decreases. At point 'I' the secondary overfills its Roche lobe
again, so that the supercritical disc is formed. 
At point 'J'
the secondary evolves into thermally pulsing AGB-star. 
At point 'K' the secondary leaves the AGB to become a hot white dwarf.
To summarize, between  points 'B'
and 'E' and from 'H' to 'K'  the considered system appears as a SyXB.
Note that the torque reversal from spin-down to spin-up
occurred close to point 'C' when the mass loss rate from the 
red giant gradually increased, in accordance with Eq. \ref{eq1}.

\section{Results and conclusion}

We have used the treatment of 
quasi-spherical wind accretion considered in \cite{Shakura_ea10} to calculate the 
properties of a population of SyXB in the Galaxy. The resulting population of SyXBs
on the $P_s-L_x$ plane is shown in Fig. \ref{fig:psma}, in which known SyXBs from 
Table \ref{tab:syxb} are also plotted by open circles. Several population synthesis models
(with different kick velocities of the nascent NS and stellar wind parameters)
were calculated (see \cite{Lu_ea10} for more detail). The main results are:

1.  The galactic birthrate of all SyXBs is $\sim (4 -   9)\times
10^{-5}$ yr$^{-1}$. The total number of galactic SyXBs is between $\sim 5000$ and
$\sim 10000$. The birthrate  of SyXB mostly depends on the kick velocity
dispersion imparted to nascent NS in core-collapse supernovae. 
However, this parameter affects the population of SyXB by a factor of about 2.

2. 70\% --- 80\%  of  NSs in SyXB are formed in
core-collapse supernovae, 10\% --- 20\%  are formed in electron-capture supernovae.  
Less than $\sim$ 1\% of SyXBs experience super-Eddington accretion.

3. The galactic number of SyXBs with bright (the absolute stellar magnitude
$M_{v}<1$) low-mass red-giant companion are from $\sim$ 40
to 120, and their birthrate is $\sim (5-8)\times
10^{-5}$~yr$^{-1}$. According to our model,
among known SyXBs, Sct X-1 and IRXS J180431.1-273932 are wind-fed
accretors. GX 1+4 lies in the transition from the wind-fed SyXBs to
SyXBs in which the giants overflow their Roche lobe.  
The model successfully reproduces very long NS spins 
(such as in IGR J16358-4724 and 4U 1954+31) without the need to invoke
very strong magnetic fields.

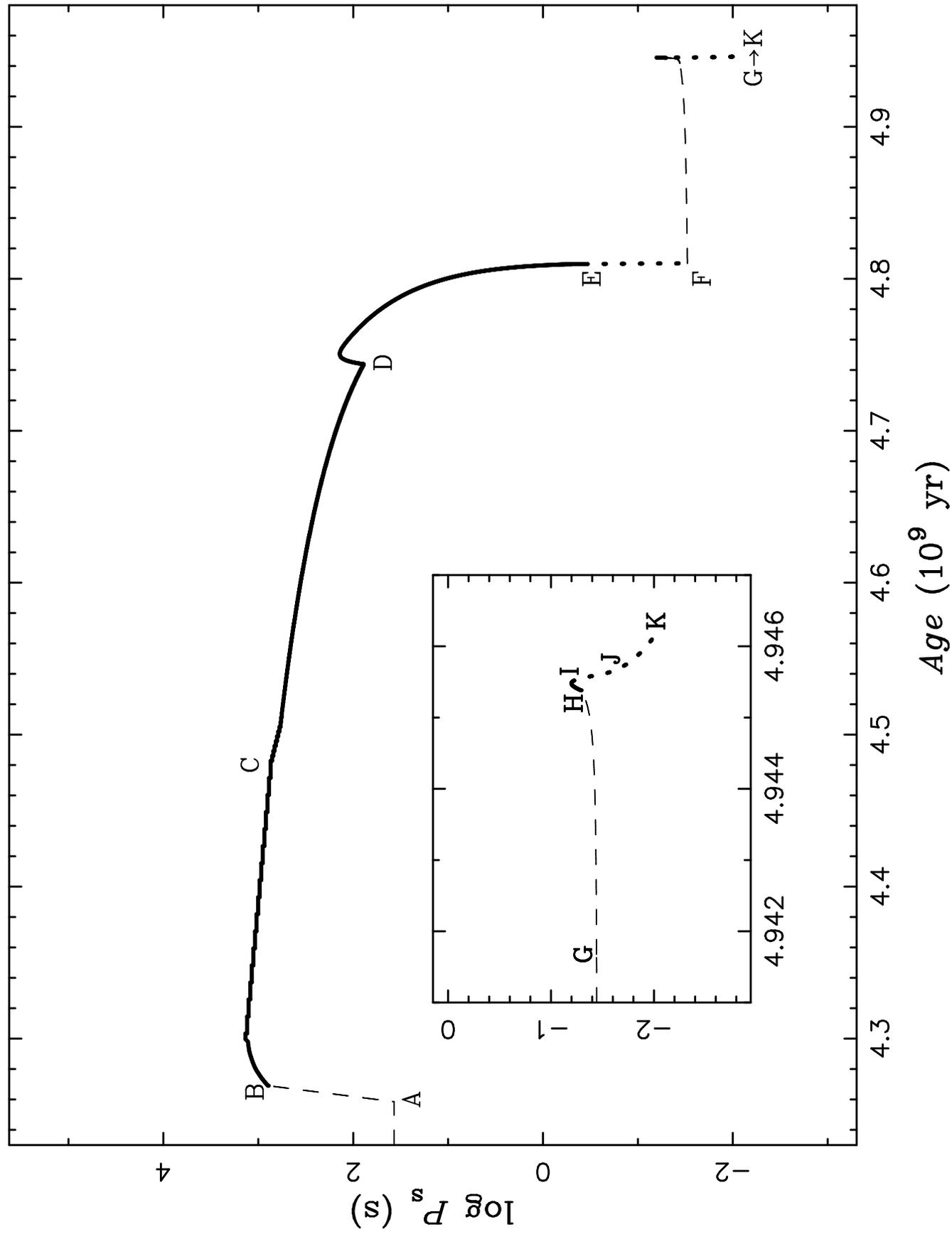
\end{document}